\documentclass[pdflatex,sn-nature]{sn-jnl}

\usepackage{graphicx}%
\usepackage{multirow}%
\usepackage{amsmath,amssymb,amsfonts}%
\usepackage{amsthm}%
\usepackage{mathrsfs}%
\usepackage[title]{appendix}%
\usepackage{xcolor}%
\usepackage{textcomp}%
\usepackage{manyfoot}%
\usepackage{booktabs}%
\usepackage{algorithm}%
\usepackage{algorithmicx}%
\usepackage{algpseudocode}%
\usepackage{listings}%
\usepackage{lmodern}
\usepackage{physics}
\usepackage{xfrac}
\usepackage[capitalise]{cleveref}
\usepackage[nolist,nohyperlinks]{acronym} 
\usepackage{lineno}      

\theoremstyle{thmstyleone}%
\theoremstyle{thmstyletwo}%

\theoremstyle{thmstylethree}%

\begin{document}

\acrodef{sw}[SW]{spin wave}
\acrodef{bec}[BEC]{Bose--Einstein condensate}
\acrodef{yig}[YIG]{yttrium iron garnet}
\acrodef{lo}[LO]{local oscillator}
\acrodef{sa}[SA]{spectrum analyzer}
\acrodef{sg}[SG]{signal generator}
\acrodef{bv}[BVSWM]{Backward--Volume Spin Wave Mode}
\acrodef{fv}[FVSWM]{Forward--Volume Spin Wave Mode}

\title{Spontaneous Emergence of Phase Coherence in a quasiparticle Bose--Einstein Condensate}
\author{Malte Koster\textsuperscript{1}}
\author{Matthias R. Schweizer\textsuperscript{1}}
\author{Timo Noack\textsuperscript{1}}
\author{Vitaliy I. Vasyuchka\textsuperscript{1}}
\author{Dmytro A. Bozhko\textsuperscript{2}}
\author{Burkard Hillebrands\textsuperscript{1}}
\author{Mathias Weiler\textsuperscript{1}}
\author{Alexander A. Serga\textsuperscript{1}}
\author{Georg von Freymann\textsuperscript{1,3}}

\affil{\textsuperscript{1}Department of Physics and State Research Center OPTIMAS, RPTU University Kaiserslautern-Landau, 67663 Kaiserslautern, Germany}
\affil{\textsuperscript{2}Center for Magnetism and Magnetic Nanostructures, Department of Physics and Energy Science, University of Colorado Colorado Springs, Colorado Springs, CO, 80918, USA}
\affil{\textsuperscript{3}Fraunhofer Institute for Industrial Mathematics ITWM, 67663 Kaiserslautern, Germany}

\abstract{
Since their prediction by Einstein at the dawn of quantum mechanics \cite{Einstein1932}, Bose--Einstein condensates (BECs), owing to their property to show quantum phenomena on macroscopic scales, are drawing increasing attention across various fields in physics. 
They are the subject of many fascinating observations in various physical systems, from liquid helium \cite{Penrose1956} to diluted atomic gases \cite{Anderson1995, Davis1995}.
In addition to real particles like atoms and composite bosons such as Cooper pairs \cite{Leggett2006} or excitons \cite{Snoke1990}, this phenomenon is also observed in gases of quasiparticles such as polaritons \cite{Balili2007} and magnons -- quanta of spin-wave excitations in magnetic media \cite{Demokritov2006}.
The fundamental property of the BEC state is its coherence \cite{Snoke2006, Rezende2009}, which is represented by a precisely defined phase of the corresponding wave function, which arises spontaneously and encompasses all particles gathered at the bottom of their spectrum. 
Until now, the BEC phase was only revealed in phenomena depending on the spatial phase difference, such as interference \cite{Andrews1997, NowikBoltyk2012}, second order coherence \cite{Deng2002}, and macroscopic BEC motions---supercurrents \cite{Bozhko2016}, superfluidity \cite{Snoke2002nature, Butov2002, Pitaevskii2016, Maekinen2024} and Josephson oscillations \cite{Pitaevskii2016, Autti2020, Kreil2021}.
Here, we present a method for the direct time-domain measurement of the magnon BEC coherent state phase relative to an outside reference signal.
We report the emergence of spontaneous coherence from a freely evolving magnon gas, which manifests as the condensation of magnons into a uniform precession state with minimal energy and a well-defined phase.
These findings confirm all postulated fundamental properties of quasiparticle condensates \cite{Snoke2006}, provide access to a new degree of freedom in such systems, and open up the possibility of information processing using microwave-frequency magnon BECs \cite{Mohseni2022}.
}


\maketitle

\section{Main}\label{intr}

The spontaneous emergence of coherence \cite{Snoke2006} is a fundamental property of \acfp{bec} \cite{Einstein1932, Penrose1956, Anderson1995, Davis1995, Leggett2006, Klaers2010}, including quasiparticle condensates \cite{Balili2007, Snoke2002, Snoke2006, Snoke2013, Snoke1990, Frohlich1968}.
In the case of magnon condensates \cite{Demokritov2006, Lvov2024, Rezende2009}, coherence is manifested in the observation of phenomena such as quantized vorticity \cite{NowikBoltyk2012}, supercurrents \cite{Bozhko2016, Schweizer2024, schweizerConfinementBoseEinstein2022}, Bogoliubov waves \cite{Bozhko2019}, or Josephson oscillations \cite{Kreil2021}.
Even though these methods provide conclusive evidence of coherence, it has been impossible to observe its spontaneous emergence. These previous observations of coherence could thus not rule out the possibility of coherent properties being introduced by external variables, such as the coherent pumping of quasiparticles into the spin system \cite{Snoke2013, Kreil2019, Demokritov2022}.
However, the formation of a random phase in concurrence with a spontaneous increase of coherence is an inherent and necessary property of a \ac{bec}, and provides evidence of the condensation process.
Since the quantum phase is not directly observable, magnonics provides a special opportunity for these measurements \cite{Pirro2021, Rezende2009}.
The magnon \ac{bec} can also be represented as a semi-classical macroscopic object, and, therefore, it is possible to determine the phase of its coherent state with respect to an external reference.
The concept of quantum-classical correspondence allows to draw conclusions about the coherence of the quantum mechanical phase from the coherence of the classical phase.
By utilizing this correspondence, we are able to observe the spontaneous establishment of \ac{bec} phases. These phases are independent of each other as well as the external reference signal and pump phases. The \ac{bec} phase is randomly formed in every repetition of the experiment.
In our work, we present a novel, electronic approach to both excite and measure a magnon \ac{bec} in a single setup, and furthermore have the ability to directly measure the evolution of the coherent state phase of the condensate as well as its amplitude. 
In turn, this enables us to directly observe the spontaneous emergence of coherence as well as the randomness of the phase in the \ac{bec}.
\begin{figure*}
    \centering
    \includegraphics[width=\textwidth]{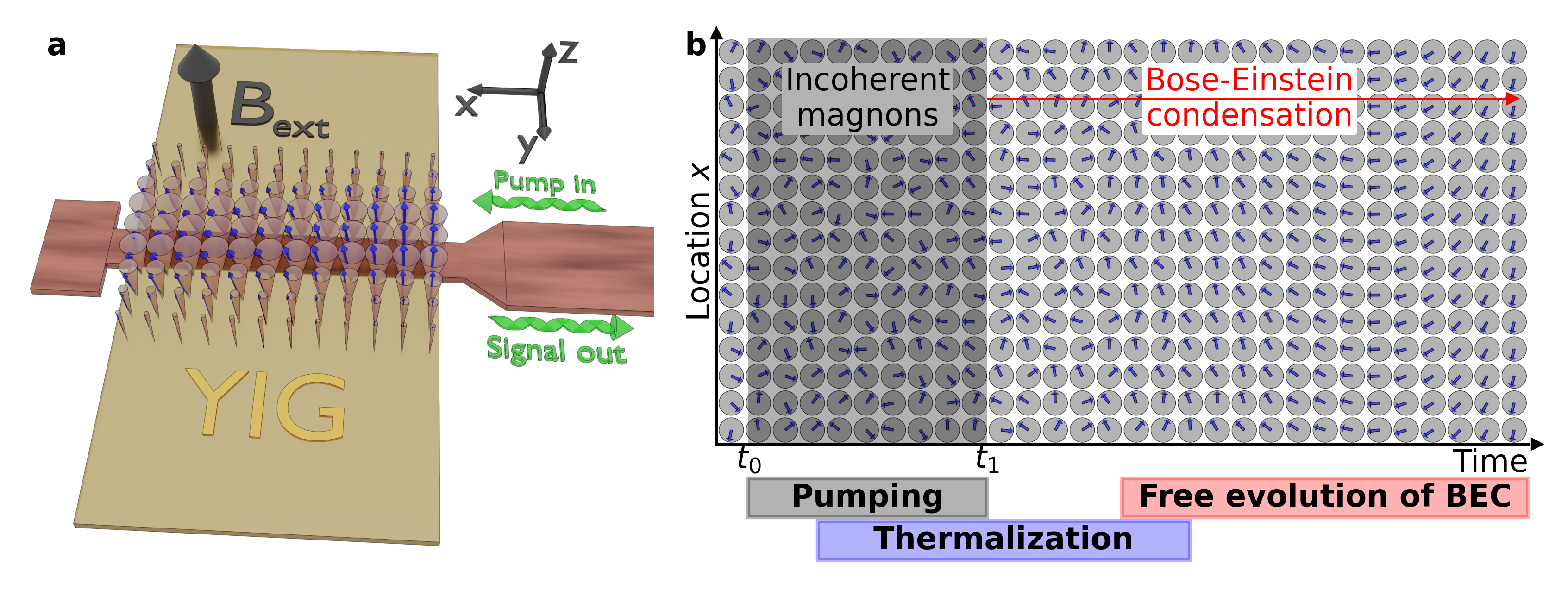}
    \caption{Sketch of the measurement setup and the magnon system dynamics. a) A microstrip antenna on an out-of-plane magnetized \ac{yig} film is connected to a microwave pump source. The alternating magnetic field of the microstrip is coupled to the ordered \ac{yig} spin system and excites magnons parametrically at half of the pumping frequency.
    The pumped magnons thermalise and undergo a transition to the \ac{bec} state. The resulting precession of the magnetization induces a signal in the antenna, which is then further amplified and detected by a phase-sensitive measurement setup.
    b) Schematic view of the spatio-temporal behaviour of the macrospin system. The different stages of the condensation process are indicated. During the pumping, coherent high-energy magnons are introduced into the system. As the density of magnons increases, thermalisation starts and the system evolves to an incoherent magnon gas state, which then transitions towards the ordered \ac{bec} state. Afterwards, only the \ac{bec} is detected, at which point the magnon signal becomes fully coherent.}
    \label{fig:fig1}
\end{figure*}

\subsection{Magnon BEC properties}\label{theo}
Magnons can be described as the delocalized quantum excitations of an ordered ensemble of spins. 
In its simplest form, this means the flipping of a single spin in an otherwise magnetically perfectly ordered material. 
Since a single flipped spin is not an eigenstate of the system, it is delocalized in the form of a Bloch state \cite{Rezende2020}.
For high occupation numbers, this leads to a macroscopic precession of the magnetization.
As quasiparticles, magnons are bosonic \cite{Holstein1940} and thus can undergo Bose--Einstein condensation. 

Magnons in magnetically ordered materials possess several defining properties, one of which is their strong anisotropy. 
In films, the magnon dispersion relation strongly depends on the orientation of the bias magnetic field, relative to the wave vector $k$ of a magnon. 
This leads to considerable deviations in the behaviour for different sample orientations. 
In contrast to most previous works on magnonic \ac{bec}, which utilized an in-plane magnetized film \cite{Demokritov2006,Demidov2008} and showed Bose--Einstein condensation in the \ac{bv} geometry at wave number $k\ne0$, we choose an out-of-plane magnetization (i.e. \ac{fv}) geometry and thus we are observing condensation at $k\approx0$. 
For further information on the experimental geometry, see the Methods section.
 
The condition for the formation of a \ac{bec} can be derived from the occupation of the Bose--Einstein distribution of a gas of bosonic particles or quasiparticles \cite{Pethick2008}
\begin{equation}
	\label{eq:ideal_be_particle_number}
	N_\mathrm{g}(T, \mu)= \int_0^\infty\frac{D(\varepsilon)}{e^{(\varepsilon-\mu)/k_{\mathrm{B}}T}-1}\dd\varepsilon,
\end{equation}
where $N_\mathrm{g}$ denotes the total number of particles in the magnon gas, $D(\varepsilon)$ is the density of states at energy $\varepsilon$, $\mu$ the chemical potential, $T$ the temperature, and $k_\mathrm{B}$ the Boltzmann constant.
For quasiparticles, it is possible to increase the number of particles in the system by external pumping \cite{Serga2014, Deng2002, Snoke2002nature}, so that,
whenever the number of particles in the system exceeds the maximum of $N_{\mathrm{g}}(T_0, \mu=\epsilon_0)$ at a finite temperature $T_0$, the excess magnons accumulate in the lowest available energy state, i.e., in the ground state \cite{Einstein1932, Lvov2024}, resulting in a thermodynamic phase transition.
In contrast to \acp{bec} in atoms, the quasiparticle (e.g., magnon) density can be largely increased, and a \ac{bec} can be observed even at room temperature. \cite{Demokritov2006, Schneider2020}

\begin{figure*}
    \centering
    \includegraphics[width=\textwidth]{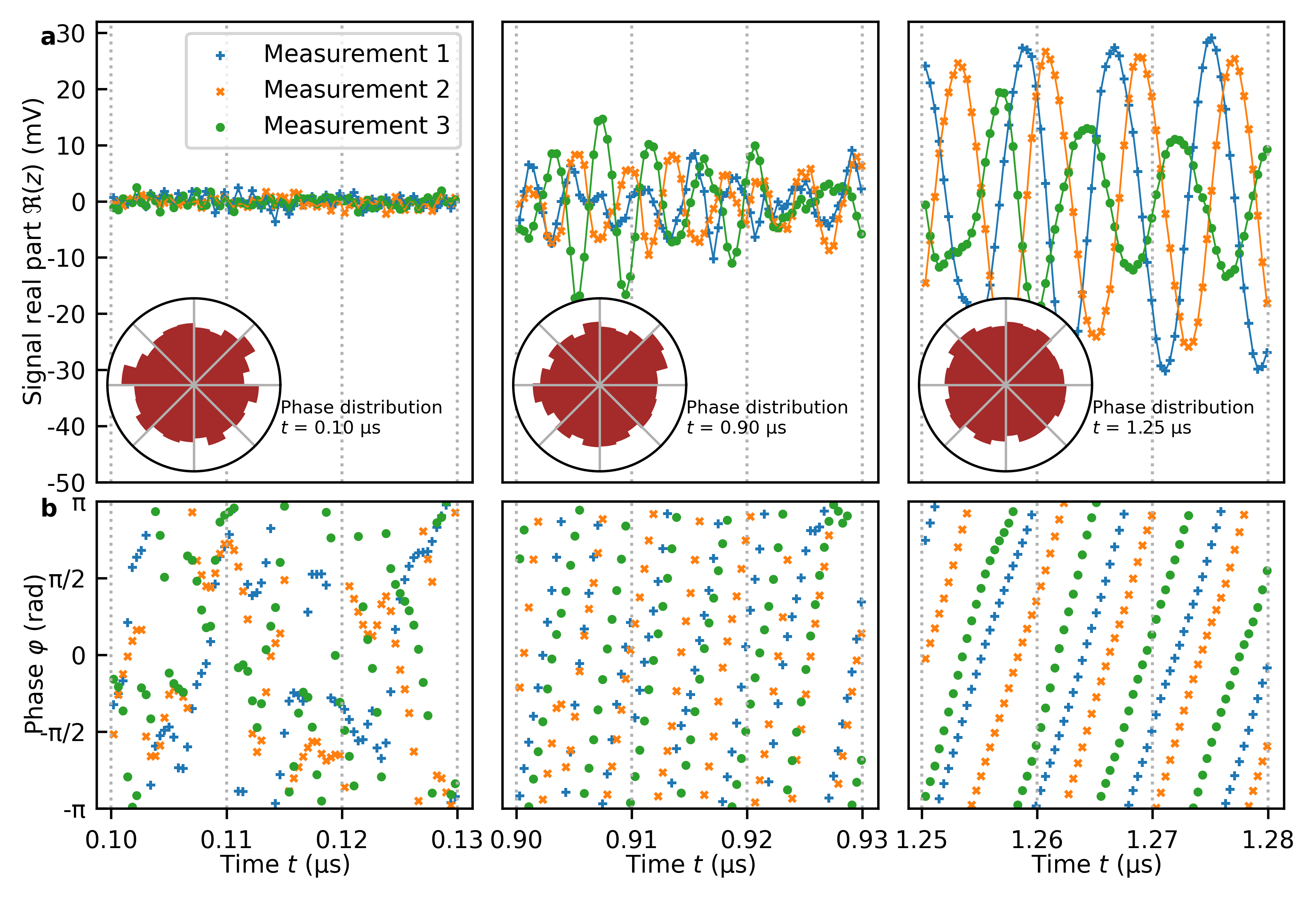}
    \caption{Time evolution of the signal. a) The real component and phase of the measured complex signal at different time slices at a pumping power of 23\,dBm. For each time slice, three randomly selected measurements are depicted. Lines are a guide to the eye. The system is pumped from $t_\mathrm{on} = 0\,\mathrm{\mu s}$ to $t_\mathrm{off} = 1\,\mathrm{\mu s}$. The pump signal is filtered out, and only oscillations near the bottom of the spectrum are measured.  The inset shows angular histograms of the phase distribution of the wave function over 1000 measurements at the beginning of each frame. Amplitude denotes the percentage of measurements in this cluster. The distribution shows no apparent clustering and is uniform, indicating randomness of the formed BEC phase in each realization of the experiment. This can be seen as a strong indication that the observed coherence is a result of true \ac{bec} and not an induced coherence by external factors \cite{Snoke2006}. b) The complex phase of the signal is analysed at the same time slices as in a). No coherent phase evolution can be observed at the beginning of the pumping pulse. As the system undergoes condensation, the phase starts to show more and more coherent behaviour. In the \ac{bec} state, a uniform phase evolution of the signal is observed.} 
    \label{fig:fig2}
\end{figure*}

In the \ac{bec} state, the ground state forms a coherent ensemble of bosons, whose behaviour is described by a single collective wave function.
A crucial property of this state is the spontaneous emergence of coherence, which means that the initial phase state of the condensate must be independent of the external pumping \cite{Snoke2006, Rezende2009}. As a consequence, in a sequence of individual single-shot experiments, the BEC generated in each of the realizations establishes a random phase.

\subsection{Phase evolution of the \ac{bec}}\label{results}

In our experiments, we observe the time evolution of a parametrically pumped gas of magnons. Both pumping and detection of the magnetic response are achieved by a microstrip antenna, inductively coupled to a \ac{yig} film (see \cref{methods} for details). The sample is pumped for 1\,µs with varying microwave powers, and the response of the system is recorded over the whole duration of pumping, as well as during the free evolution of the magnonic system up to 2.5\,µs after the pumping is switched off. For a complete sketch of the setup, see Extended Data \cref{fig:figext1}. It is important to note that this approach enables us to perform single-shot measurements without the need for integration over consecutive measurements. The detection path is not sensitive to the parametrically introduced magnons, but only to the thermal gas near the bottom of the spectrum and at the bottom itself.

The time evolution of a measurement can be divided into three sections: (i) The pumping, during which new magnons are introduced into the system by parametric pumping. Depending on the power and coupling efficiency, the rate of magnon injection can sufficiently exceed the decay in the system, and thus, the critical density for condensation can be reached. As the magnon density reaches the critical threshold, (ii) the thermalisation section begins, and the excess magnon gas thermalises into the coherent condensate state. Finally, after the pump is switched off and the condensate is fully formed, we can observe (iii) the free evolution of the magnon \ac{bec}. These sections are schematically illustrated in \cref{fig:fig1}.
 
Using the directly measured signal, we are able to determine the phase of the magnetization precession at arbitrary points in time during and after pumping. The evolution of the phase is a clear measure of coherence. A non-coherent signal exhibits a random phase trace, whereas a coherent signal manifests itself with a linear phase evolution. The time dependencies of the signal before and after condensation for three independent measurements at 23\,dBm are shown in \cref{fig:fig2}. In our setup, we record a signal which is in phase to an external reference \ac{lo}, as well as a 90 degree shifted one, which will in turn be interpreted as the real and imaginary parts of a complex function, i.e. in-phase $I(t)$ and quadrature $Q(t)$ components (see \cref{sec:experiment}). Thus, the phase of the oscillation can be determined as the complex angle between the real and imaginary parts, as shown in \cref{fig:fig2}b). The signal and its phase reach a high degree of coherence only after condensation. Further analysis of the phase shows uniform phase development, indicating a coherent condensate, and also the spontaneity of the initial phase in every experiment. 
This analysis is presented in \cref{methods:da}. The inlays in \cref{fig:fig2} show histograms of the phase at $t_0$ of 1000 independent measurements. At the beginning of the pumping sequence, the signal amplitude, which is proportional to the square root of the magnon density, is low, and no coherent behaviour is apparent. As more and more magnons are injected into the system, the signal starts to show some periodic and apparently coherent traits. After the pump is switched off, condensation occurs, and an increase in the population of the ground state is observed. At the same time, the coherence of the signal increases.
Since all generation and detection equipment in the experimental setup is phase-locked, and the signal generator primed to start pumping with a constant phase-offset throughout the whole measurement, all externally induced factors are stable within a few degrees of phase shift, and the uniform phase distribution must be an intrinsic property of the \ac{bec}. This leads to the conclusion that the emergence of a stable phase evolution is indeed spontaneous and not induced by external factors.

\begin{figure*}
    \centering
    \includegraphics[width=\textwidth]{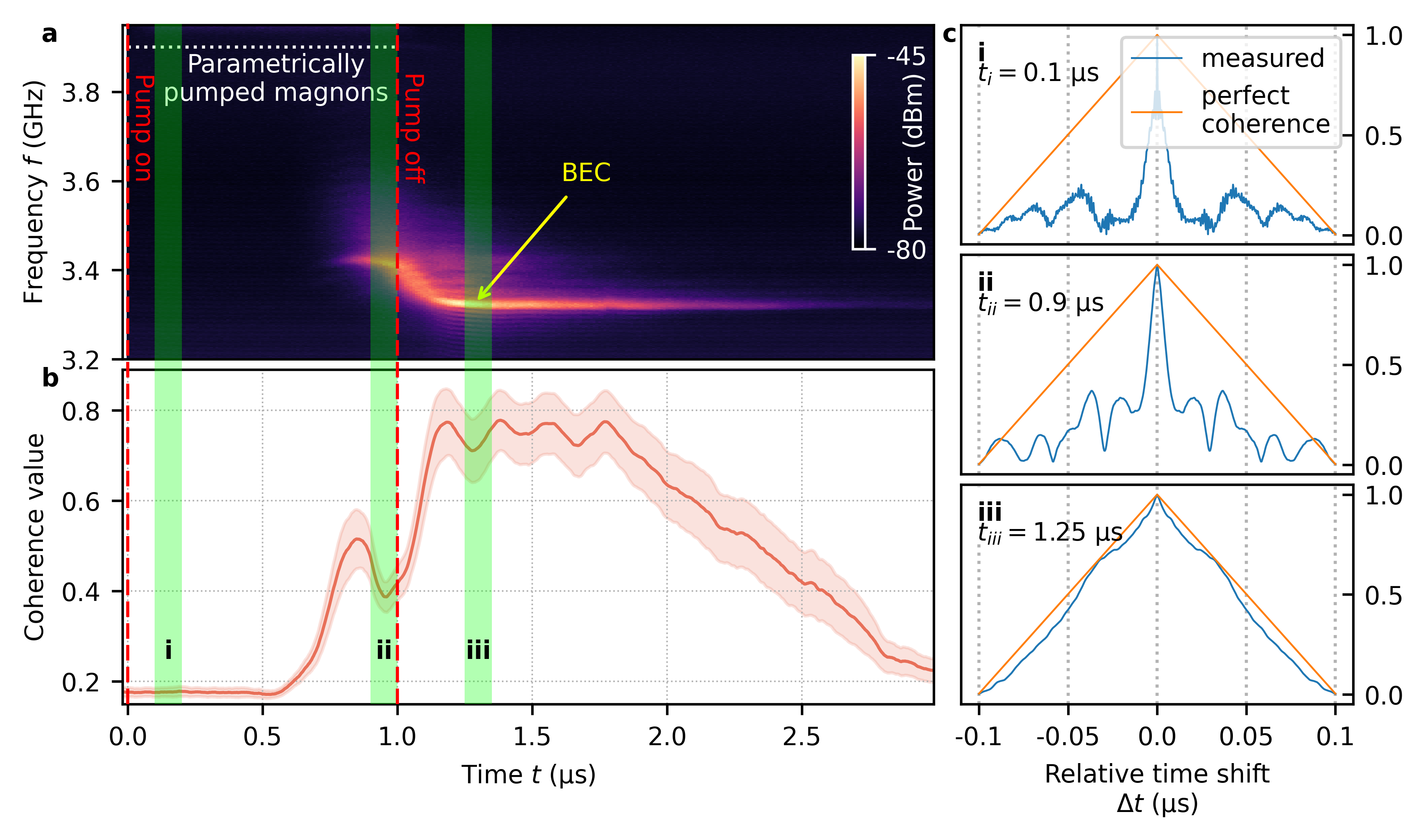}  
    \caption{Emergence of coherence at a pumping power of 23\,dBm. a) Spectrogram of the signal during the measurement. 
    The condensation process can be observed after the density of the magnon gas reaches the critical value for condensation, as indicated by the increase in signal amplitude. The downwards shift of the frequency is an effect of the decrease in the overall number of magnons after the pump is switched off and the corresponding increase in net magnetization. 
    b) The evolution of coherence as a measure of the quotient of the integrals of the autocorrelation of the signal and the ideal monochromatic signal. 
    A value of one indicates a perfectly coherent signal, and zero denotes perfect white noise. The given measurements are an average of 1000 measurements, with the shaded area denoting the standard deviation. The emergence of coherence can clearly be observed.
    After the pump is switched off and thermalisation is finished, the \ac{bec} is fully formed and peak coherence is reached.
    c) Indicated are the absolute values of the autocorrelation function of the measured signal at different time-windows, compared to that of a fully coherent $f(t)=e^{i\omega_0 t}$ signal of the same frequency. The three slices are $100\,\mathrm{ns}$ wide and are located at $t_i=0.1\,\mathrm{\mu s}$, $t_{ii}=0.9\,\mathrm{\mu s}$ and $t_{iii}=1.25\,\mathrm{\mu s}$. 
The ratio of the integral of the measured autocorrelation is used to calculate coherence values seen in b).}
    \label{fig:fig3} 
\end{figure*}

\subsection{Emergence of coherence}

As a measure of coherence, we used the integral of the normalized autocorrelation of the direct signal over a 100\,ns boxcar window, shifted over the entire measurement interval. Comparing this to the theoretically ideal autocorrelation (i.e., the autocorrelation of a perfect $f(t)=e^{i\omega t}$ pulse with a boxcar envelope) allows us to generate a dimensionless measure of coherence -- the coherence value, where 0 denotes perfect white noise and 1 denotes a perfectly coherent signal. It is important to note that this value is calculated for the centre of each reference frame, and thus has the same time resolution as the original signal.

\cref{fig:fig3} shows the evolution of frequency and coherence throughout the measurement. At the beginning of the pumping, the signal is not registered at all, as the injected parametric magnons are not yet thermalised and are located outside of the detection frequency band. As the thermalisation process evolves, an increase in coherence can be observed. After the pumping is completed, the frequency of the ground state drifts towards lower frequencies, causing a temporary reduction of coherence. This phenomenon can be attributed to the high density of magnons during the pumping section and the resulting nonlinear shift due to magnon-magnon interactions \cite{Verba2019}. Once the magnon gas has completely thermalised and the \ac{bec} has absorbed the excess of the hot, thermal magnons, the coherence value reaches its final value. The gradual decrease in coherence, late after the pump is turned off, can be understood by the decrease in amplitude and thus the deterioration of the signal-to-noise ratio.
\begin{figure*}
    \centering
    \includegraphics[width=0.5\textwidth]{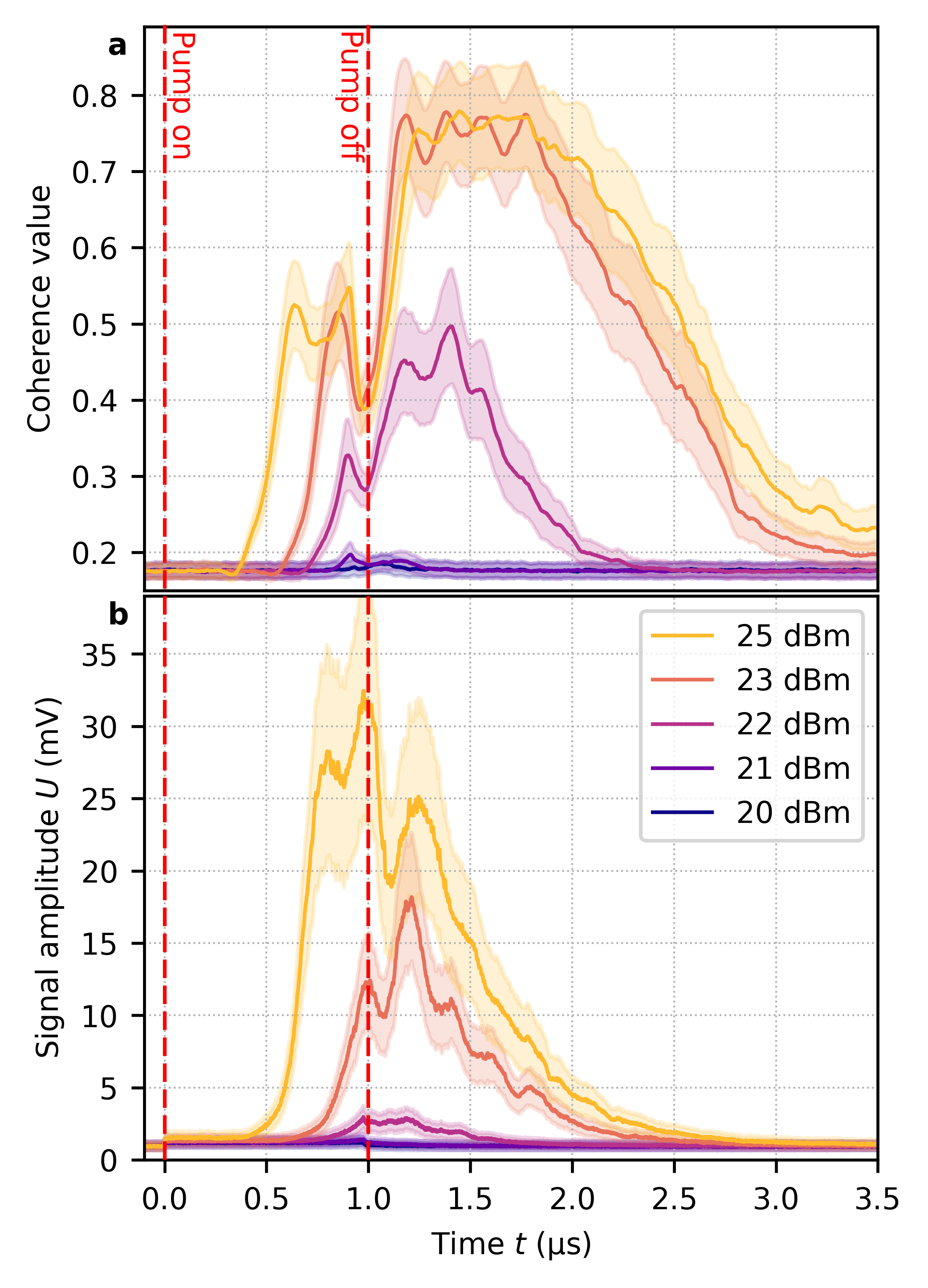}
    \caption{Power dependence of coherence value and amplitude. By changing the pump power, a threshold for condensation can be obtained. At $20\,\mathrm{dBm}$, there is no increase in coherence. At $21\,\mathrm{dBm}$, an increase in coherence is observed, around $t=1\,\mathrm{\mu s}$. Due to the low amplitude and the resulting low SNR, the increase of the coherence value is not very pronounced. At $22\,\mathrm{dBm}$, a strong increase in coherence can be observed, although the amplitude of the signal is still very low. This indicates a condensation threshold of $21\,\mathrm{dBm}$ in this particular setup. Increasing the pump power above $23\,\mathrm{dBm}$ does not lead to higher coherence, although a higher amplitude can be observed. Remarkably, the coherence of the \ac{bec} does not depend on the absolute amplitude of the signal. At higher powers, high levels of coherence can be seen even after the amplitude significantly decreases. This is a strong indication that the \ac{bec} doesn't lose its coherent properties during decay.}
    \label{fig:fig4}
\end{figure*}
A comparison of the coherence value with the amplitude of the signal also shows that the \ac{bec} is phase-stable until the signal vanishes due to the finite lifetime of the magnons. In other words, the \ac{bec} does not undergo dephasing, but the magnonic damping lowers the amplitude of the oscillation until the signal becomes indistinguishable from noise. At this point, the signal itself is no longer detectable, although the correlation analysis (i.e., the coherence value) still shows a clearly detectable coherence. This can be seen in \cref{fig:fig4}. At $t=3.5\,\mathrm{\mu s}$ the amplitude of the signal has vanished below the noise threshold, but a clear remanence in coherence can be observed. 

\section{Discussion}\label{conc}
While the existence of \acp{bec} in a variety of systems was demonstrated by former studies (see e.g. \cite{Maekinen2024}), no direct proof of randomness in the established condensate phase has yet been provided.
Our system provides access to the coherent state phase, which is in our semi-classical system equivalent to the quantum mechanical phase, allowing to directly probe the coherence of the BEC state.
In this work, we demonstrate the emergence of phase coherence during the formation of a \ac{bec} of quasiparticles, applying a novel phase-sensitive electro-magnetic measurement technique to magnons --- the quanta of spin waves.  
With this approach, we succeeded in exploiting the quantum-classical correspondence in magnonic systems to overcome the fundamental hurdles in the study of coherent states phase evolution, thus opening up new perspectives in the field of experimental research of condensates.
For the first time, we are able to directly measure the spontaneous emergence of coherence of magnon condensate in a solid-state system and directly show independence of the established condensate phase from any external sources, which has long been required as a necessary feature to prove the formation of a true BEC \cite{Snoke2006}.

\section{Methods}\label{methods}

\subsection{Experimental Setup}\label{methods:exp}
\begin{figure*}[h]
    \centering
    \includegraphics[width=\textwidth]{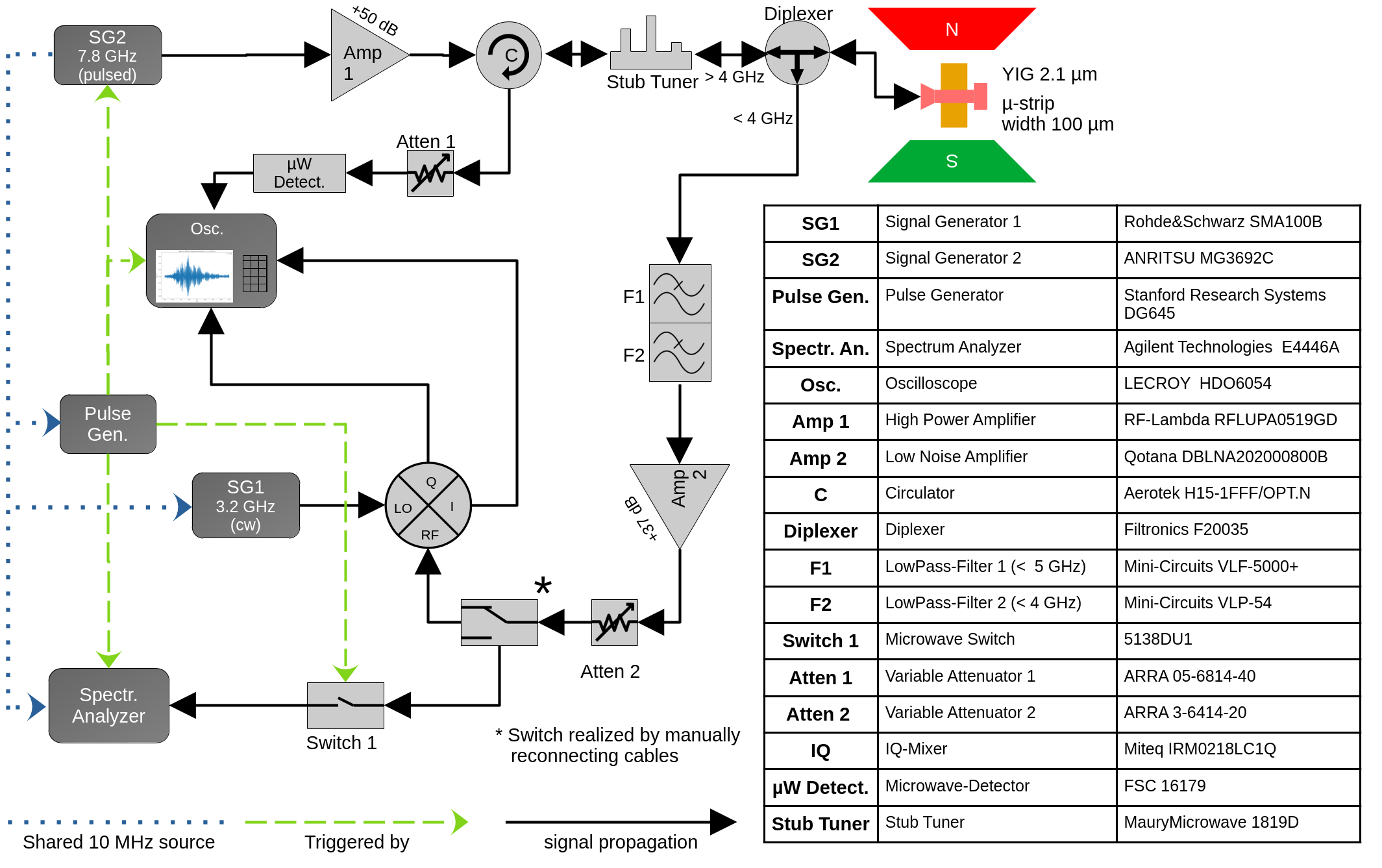}
    \caption{Complete schematics of the setup. The pump frequency is generated at the signal generator SG2. After amplification, the pumping frequency is led through a circulator, stub resonator, and diplexer to obtain the required signal for tuning the setup to the pumping frequency. The reflected part of the signal with a frequency close to the band bottom is then further filtered and amplified, before being either detected by the spectrum analyser or the IQ-mixer and oscilloscope. To avoid unnecessary distortion due to further components, the switch was realized by manually reconnecting the cables.
    All components that rely on an internal oscillator are synced by a shared 10\,MHz source, to avoid relative frequency- or phase-drift. The oscilloscope was excluded due to technical difficulties, but phase stability of the setup as a whole was confirmed in preliminary measurements.}
    \label{fig:figext1}
\end{figure*}

\begin{figure*}[h!]
    \centering
    \includegraphics[width=\textwidth]{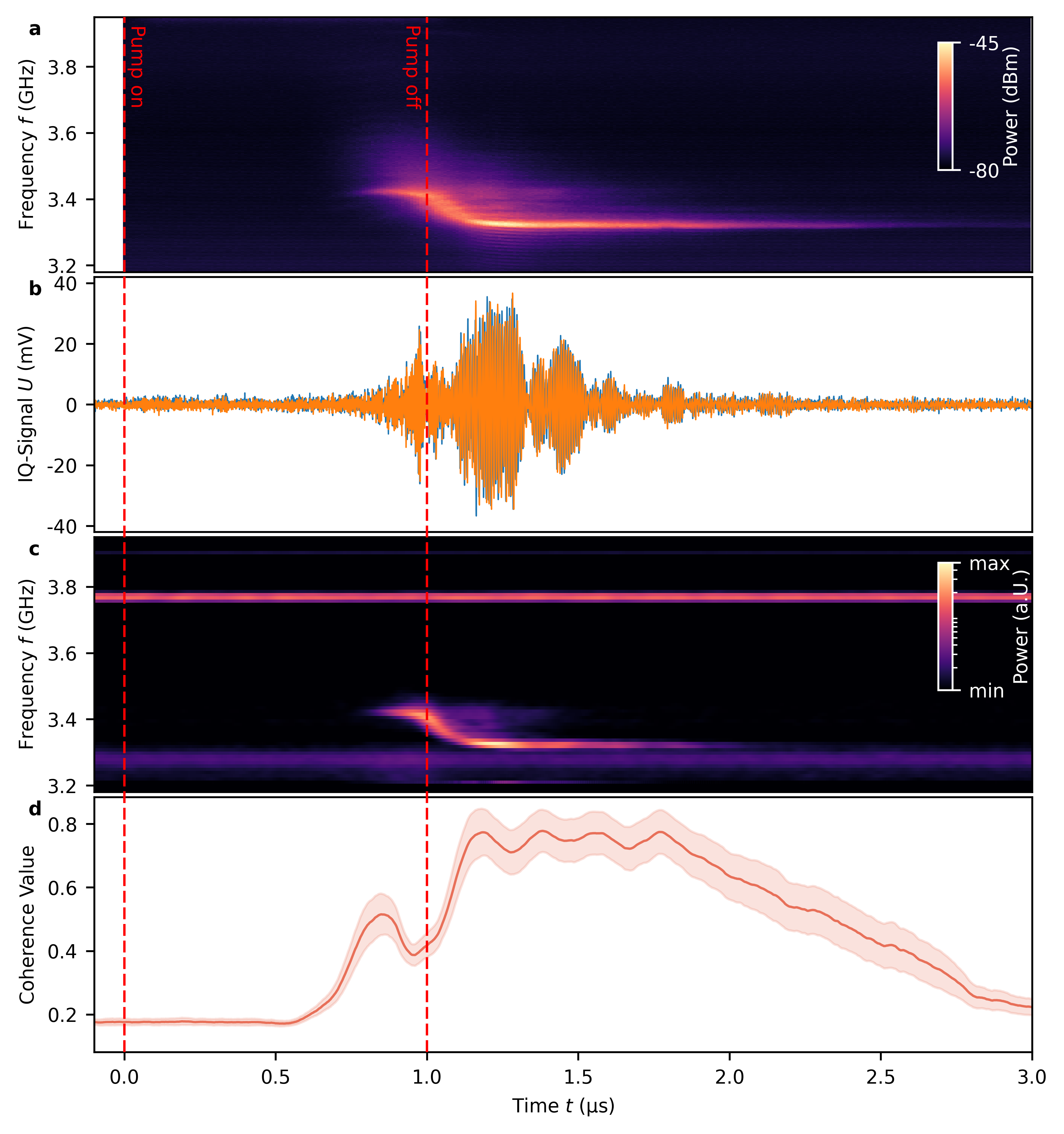}
    \caption{Comparison of different measurement techniques at a pumping power of 23\,dBm. a) Spectrogram taken by the spectrum analyser by integration. b) Exemplary raw I and Q data of a single measurement, taken by the oscilloscope. One can clearly distinguish the pumping interval, as well as the sudden increase in amplitude, as the high-energy gas condenses into the \ac{bec}. c) The spectrogram obtained by using a Fourier-Transform with a window length of 100\,ns on the complex signal, calculated from the IQ-data. The spectrogram is calculated for 1000 distinct measurements and averaged in the end. The general form of the signal matches the data obtained in a), confirming the results. Nonetheless, artifacts of the IQ-technique in the form of a constant band at 3.77\,GHz and a shadow between 3.2\,GHz and 3.3\,GHz are clearly visible. d) The evolution of the coherence value for comparison.}
    \label{fig:figext2}
\end{figure*}

The coherence measurements were performed by pumping a 2.1\,µm thick film of \ac{yig} at 7.8\,GHz with varying powers through a microstrip antenna for 1\,µs at an external magnetic field of 281\,mT. The reflected signal was separated using a frequency diplexer, filtered by a 5\,GHz low-pass filter to suppress all remains of the strong pumping, and amplified by a low-noise amplifier (Qotana DBLNA202000800B) before being fed into the RF port of an IQ mixer (Miteq IRM0218LC1Q). The IQ mixer was supplied with a 3.2\,GHz local oscillator frequency, and the I and Q outputs were routed to an oscilloscope (Teledyne LeCroy HDO6054). A more detailed description of the working mechanisms of the IQ mixer setup can be found in the supplementary material \cref{sec:experiment}. The detailed setup is depicted in the Extended Data \cref{fig:figext1}.
For the acquisition of the displayed spectra, the IQ mixer and oscilloscope were replaced by a microwave switch (Miteq S138BDU1) and a spectrum analyser (Agilent E4446A). The consistency of the data obtained by the spectrum analyser and the IQ-mixer has been validated, as shown in the Extended Data \cref{fig:figext2}

\subsection{Data Analysis}\label{methods:da}
The method chosen to determine the coherence value was the autocorrelation of the normalized complex signal over a 100\,ns wide window, using the NumPy Python package~\cite{numpy}. Since the signal $z(t) = I(t) + iQ(t)$ is complex, this leads to a complex autocorrelation function. However, only the absolute value of the autocorrelation function is relevant for determining coherence. To calculate the coherence value at a given time $t_0$, the absolute value of the autocorrelation function of $z$ is integrated, and the resulting value is normalized by dividing it by the integral of the absolute value of the autocorrelation function of a perfectly coherent dummy signal $f(t)=e^{i\omega_0 t}$. Thus, the normalized coherence value is obtained. 

\section{End Notes}
Supplementary Information is available for this paper.

\section{Acknowledgments}
This research was funded by the Deutsche Forschungsgemeinschaft (DFG, German Research Foundation) in the framework of TRR 173 -- Grant No. 268565370 Spin+X (Projects B04 and B13), as well as by grant DMR-2338060 from the National Science Foundation of the United States.\\
The authors thank Yaroslav Tserkovnyak and Vasyl Tyberkevych for the valuable discussions on the details of the magnonic condensation process and coherency questions.

\section{Author Contributions}
T.N., V.I.V., and A.A.S. conducted preliminary studies of BEC in the uniform precession state. The phase-resolved experiment was conceptualized by M.K., B.H., M.W., A.A.S., and G.v.F. with help from D.A.B. The experimental design was conducted by M.K., V.I.V., D.A.B., and M.W. with help from M.R.S. and A.A.S. The experiment was carried out by M.K. with help from V.I.V. Data analysis was performed by M.K. with help from M.R.S. The manuscript was composed by M.K., M.R.S., A.A.S., with help from M.W., V.I.V., and G.v.F. All co-authors participated in discussing the results and editing the manuscript.

\clearpage\section{Supplementary}
\subsection{Datasets and analysis}
The full dataset and algorithms used in this work are available at reasonable request. 

\subsection{BEC detection via IQ-mixing}\label{sec:experiment}
The concept of phase coherence is of central importance for the understanding of the emergent properties of the \ac{bec}. The measurement of the phase of the magnon signal, on the other hand, is hard to achieve. The two fundamental problems for gathering phase data are, first, the problem of gauge freedom if there is no external reference signal, and second, the high frequency and noise of the measured signal, which leads to poor definition of phase. Both of these challenges can be solved by utilizing an IQ-mixer setup.
Our setup consists of a pulsed microwave generator (called pump), which introduces magnons into the sample at 7.8\,GHz through a microstrip antenna. In the used out-of-plane geometry, the minimum of the dispersion relation corresponds to the $k=0$ mode, which lends itself to efficient detection by the antenna.

The pump is phase-locked to the other components in the setup to mitigate phase drift between different devices. The magnons introduced in the \ac{yig}-film then undergo Bose--Einstein Condensation, which produces a downward shift in frequency to about 3.3\,GHz. The same antenna is used for both pumping and the detection of the \ac{bec}. The \ac{bec}-signal is then directed towards a spectrum analyser as well as an IQ-Mixer. The spectrum analyser is set up to record the spectrum of the signal at different points in time during the pumping pulse as well as afterwards.

Utilizing an IQ-mixer as a heterodyne detector to detect the \ac{bec}-signal itself provides us with some unique advantages. Most importantly, this approach enables us to down convert the microwave signal to lower frequencies, while still preserving the full information of the phase with respect to an external reference signal, the so-called \acf{lo}. This \ac{lo} provides a carrier signal, on which the microwave signal is modulated. The resulting real and imaginary parts of the output form a complex signal, which leads to a trivial definition of the phase:
\begin{align}
    \varphi (t) &= \arctan \left(I(t)+i\cdot Q(t)\right)\\
    A(t) &= \abs{I(t)+i\cdot Q(t)}
\end{align}
where $\varphi(t)$ denotes the \ac{bec}-phase and $A(t)$ the amplitude of the measured signal. $A(t)$ depends on the number of magnons in the system, as well as on several experimental factors, which means it is not possible to directly draw a conclusion regarding the number of magnons in the system. Nevertheless, $A(t)$ is strongly correlated to the amplitude of the magnon signal, which means that it allows for a comparison between different measurements at identical parameters.

$\varphi(t)$, on the other hand, can be used as a direct indicator of the phase of the \ac{bec}. $I(t)$ and $Q(t)$ are collected with respect to the external \ac{lo} signal. This signal is phase locked to the pulse generator and parameters are chosen, such that the \ac{lo} remains phase stable over all measurements. The direct conclusion of this is that all deviations in the phase are, in fact, not a result of external pumping, but an intrinsically emerging property of the system at hand.
One has to note that the signal observed is, in fact, not the original signal, picked up by the antenna, but instead downshifted in frequency. Nonetheless, this has no influence on either the phase coherence or the phase randomness we observed in our data.

A downside to using an IQ-mixer for detection is the relatively narrow bandwidth. The chosen IQ-Mixer has a bandwidth of only 500\,MHz, which means it is not capable of sampling the signals corresponding to thermal magnons. To provide a full picture, a spectrum analyser was added to record the full thermal gas of magnons at different times before, during, and after the condensation. This was done to demonstrate the agreement with previous works \cite{Noack2021}.

\newpage

\renewcommand{\figurename}{Extended Data Fig.}

\newpage

\bibliography{SpontaneousCoherence}

\begin{thebibliography}{10}
\expandafter\ifx\csname url\endcsname\relax
  \def\url#1{\burl{#1}}\fi
\expandafter\ifx\csname urlprefix\endcsname\relax\def\urlprefix{URL }\fi
\providecommand{\bibinfo}[2]{#2}
\providecommand{\eprint}[2][]{\url{#2}}
\providecommand{\doi}[1]{\url{https://doi.org/#1}}
\bibcommenthead

\bibitem{Einstein1932}
\bibinfo{author}{Einstein, A.}
\newblock \emph{\bibinfo{title}{Quantentheorie des einatomigen idealen Gases.
  Zweite Abhandlung}}, \bibinfo{pages}{245--257} (\bibinfo{publisher}{John
  Wiley \& Sons, Ltd}, \bibinfo{year}{2005}).
\newblock
  \urlprefix\url{https://onlinelibrary.wiley.com/doi/abs/10.1002/3527608958.ch28}.
\newblock
  \eprint{https://onlinelibrary.wiley.com/doi/pdf/10.1002/3527608958.ch28}.

\bibitem{Penrose1956}
\bibinfo{author}{Penrose, O.} \& \bibinfo{author}{Onsager, L.}
\newblock \bibinfo{title}{{Bose--Einstein condensation and liquid helium}}.
\newblock \emph{\bibinfo{journal}{Phys. Rev.}} \textbf{\bibinfo{volume}{104}},
  \bibinfo{pages}{576--584} (\bibinfo{year}{1956}).
\newblock \urlprefix\url{http://dx.doi.org/10.1103/PhysRev.104.576}.

\bibitem{Anderson1995}
\bibinfo{author}{Anderson, M.~H.}, \bibinfo{author}{Ensher, J.~R.},
  \bibinfo{author}{Matthews, M.~R.}, \bibinfo{author}{Wieman, C.~E.} \&
  \bibinfo{author}{Cornell, E.~A.}
\newblock \bibinfo{title}{{Observation of Bose--Einstein condensation in a
  dilute atomic vapor}}.
\newblock \emph{\bibinfo{journal}{Science}} \textbf{\bibinfo{volume}{269}},
  \bibinfo{pages}{198--201} (\bibinfo{year}{1995}).
\newblock \urlprefix\url{http://dx.doi.org/10.1126/science.269.5221.198}.

\bibitem{Davis1995}
\bibinfo{author}{Davis, K.~B.} \emph{et~al.}
\newblock \bibinfo{title}{{Bose--Einstein condensation in a gas of sodium
  atoms}}.
\newblock \emph{\bibinfo{journal}{Phys. Rev. Lett.}}
  \textbf{\bibinfo{volume}{75}}, \bibinfo{pages}{3969--3973}
  (\bibinfo{year}{1995}).
\newblock \urlprefix\url{http://dx.doi.org/10.1103/PhysRevLett.75.3969}.

\bibitem{Leggett2006}
\bibinfo{author}{Leggett, A.~J.}
\newblock \emph{\bibinfo{title}{Quantum Liquids: Bose Condensation and Cooper
  Pairing in Condensed-Matter Systems}}  (\bibinfo{publisher}{Oxford University
  Press}, \bibinfo{year}{2006}).
\newblock
  \urlprefix\url{http://dx.doi.org/10.1093/acprof:oso/9780198526438.001.0001}.

\bibitem{Snoke1990}
\bibinfo{author}{Snoke, D.~W.}, \bibinfo{author}{Wolfe, J.~P.} \&
  \bibinfo{author}{Mysyrowicz, A.}
\newblock \bibinfo{title}{{Evidence for Bose--Einstein condensation of excitons
  in Cu$_2$O}}.
\newblock \emph{\bibinfo{journal}{Phys. Rev. B}} \textbf{\bibinfo{volume}{41}},
  \bibinfo{pages}{11171--11184} (\bibinfo{year}{1990}).
\newblock \urlprefix\url{http://dx.doi.org/10.1103/PhysRevB.41.11171}.

\bibitem{Balili2007}
\bibinfo{author}{Balili, R.}, \bibinfo{author}{Hartwell, V.},
  \bibinfo{author}{Snoke, D.}, \bibinfo{author}{Pfeiffer, L.} \&
  \bibinfo{author}{West, K.}
\newblock \bibinfo{title}{{Bose--Einstein condensation of microcavity
  polaritons in a trap}}.
\newblock \emph{\bibinfo{journal}{Science}} \textbf{\bibinfo{volume}{316}},
  \bibinfo{pages}{1007--1010} (\bibinfo{year}{2007}).
\newblock \urlprefix\url{http://dx.doi.org/10.1126/science.1140990}.

\bibitem{Demokritov2006}
\bibinfo{author}{Demokritov, S.~O.} \emph{et~al.}
\newblock \bibinfo{title}{{B}ose--{E}instein condensation of quasi-equilibrium
  magnons at room temperature under pumping}.
\newblock \emph{\bibinfo{journal}{Nature}} \textbf{\bibinfo{volume}{443}},
  \bibinfo{pages}{430--433} (\bibinfo{year}{2006}).
\newblock \urlprefix\url{http://dx.doi.org/10.1038/nature05117}.

\bibitem{Snoke2006}
\bibinfo{author}{Snoke, D.}
\newblock \bibinfo{title}{Coherent questions}.
\newblock \emph{\bibinfo{journal}{Nature}} \textbf{\bibinfo{volume}{443}},
  \bibinfo{pages}{403--404} (\bibinfo{year}{2006}).
\newblock \urlprefix\url{http://dx.doi.org/10.1038/443403a}.

\bibitem{Rezende2009}
\bibinfo{author}{Rezende, S.~M.}
\newblock \bibinfo{title}{Theory of coherence in {B}ose--{E}instein
  condensation phenomena in a microwave-driven interacting magnon gas}.
\newblock \emph{\bibinfo{journal}{Phys. Rev. B}} \textbf{\bibinfo{volume}{79}},
  \bibinfo{pages}{174411} (\bibinfo{year}{2009}).
\newblock \urlprefix\url{http://dx.doi.org/10.1103/PhysRevB.79.174411}.

\bibitem{Andrews1997}
\bibinfo{author}{Andrews, M.~R.} \emph{et~al.}
\newblock \bibinfo{title}{Observation of interference between two {B}ose
  condensates}.
\newblock \emph{\bibinfo{journal}{Science}} \textbf{\bibinfo{volume}{275}},
  \bibinfo{pages}{637--641} (\bibinfo{year}{1997}).
\newblock
  \urlprefix\url{https://www.science.org/doi/abs/10.1126/science.275.5300.637}.

\bibitem{NowikBoltyk2012}
\bibinfo{author}{Nowik-Boltyk, P.}, \bibinfo{author}{Dzyapko, O.},
  \bibinfo{author}{Demidov, V.~E.}, \bibinfo{author}{Berloff, N.~G.} \&
  \bibinfo{author}{Demokritov, S.~O.}
\newblock \bibinfo{title}{Spatially non-uniform ground state and quantized
  vortices in a two-component {B}ose--{E}instein condensate of magnons}.
\newblock \emph{\bibinfo{journal}{Sci. Rep.}} \textbf{\bibinfo{volume}{2}},
  \bibinfo{pages}{482} (\bibinfo{year}{2012}).
\newblock \urlprefix\url{http://dx.doi.org/10.1038/srep00482}.

\bibitem{Deng2002}
\bibinfo{author}{Deng, H.}, \bibinfo{author}{Weihs, G.},
  \bibinfo{author}{Santori, C.}, \bibinfo{author}{Bloch, J.} \&
  \bibinfo{author}{Yamamoto, Y.}
\newblock \bibinfo{title}{Condensation of semiconductor microcavity exciton
  polaritons}.
\newblock \emph{\bibinfo{journal}{Science}} \textbf{\bibinfo{volume}{298}},
  \bibinfo{pages}{199--202} (\bibinfo{year}{2002}).
\newblock
  \urlprefix\url{https://www.science.org/doi/abs/10.1126/science.1074464}.

\bibitem{Bozhko2016}
\bibinfo{author}{Bozhko, D.~A.} \emph{et~al.}
\newblock \bibinfo{title}{Supercurrent in a room-temperature {B}ose--{E}instein
  magnon condensate}.
\newblock \emph{\bibinfo{journal}{Nat. Phys.}} \textbf{\bibinfo{volume}{12}},
  \bibinfo{pages}{1057--1062} (\bibinfo{year}{2016}).
\newblock \urlprefix\url{http://dx.doi.org/10.1038/nphys3838}.

\bibitem{Snoke2002nature}
\bibinfo{author}{Snoke, D.}, \bibinfo{author}{Denev, S.}, \bibinfo{author}{Liu,
  Y.}, \bibinfo{author}{Pfeiffer, L.} \& \bibinfo{author}{West, K.}
\newblock \bibinfo{title}{Long-range transport in excitonic dark states in
  coupled quantum wells}.
\newblock \emph{\bibinfo{journal}{Nature}} \textbf{\bibinfo{volume}{418}},
  \bibinfo{pages}{754--757} (\bibinfo{year}{2002}).
\newblock \urlprefix\url{https://doi.org/10.1038/nature00940}.

\bibitem{Butov2002}
\bibinfo{author}{Butov, L.~V.}, \bibinfo{author}{Gossard, A.~C.} \&
  \bibinfo{author}{Chemla, D.~S.}
\newblock \bibinfo{title}{Macroscopically ordered state in an exciton system}.
\newblock \emph{\bibinfo{journal}{Nature}} \textbf{\bibinfo{volume}{418}},
  \bibinfo{pages}{751--754} (\bibinfo{year}{2002}).
\newblock \urlprefix\url{https://doi.org/10.1038/nature00943}.

\bibitem{Pitaevskii2016}
\bibinfo{author}{Pitaevskii, L.} \& \bibinfo{author}{Stringari, S.}
\newblock \emph{\bibinfo{title}{Bose--Einstein Condensation and Superfluidity}}
   (\bibinfo{publisher}{Oxford University Press}, \bibinfo{year}{2016}).
\newblock
  \urlprefix\url{http://dx.doi.org/10.1093/acprof:oso/9780198758884.001.0001}.

\bibitem{Maekinen2024}
\bibinfo{author}{Mäkinen, J.~T.}, \bibinfo{author}{Autti, S.} \&
  \bibinfo{author}{Eltsov, V.~B.}
\newblock \bibinfo{title}{Magnon bose–einstein condensates: From time
  crystals and quantum chromodynamics to vortex sensing and cosmology}.
\newblock \emph{\bibinfo{journal}{Applied Physics Letters}}
  \textbf{\bibinfo{volume}{124}}, \bibinfo{pages}{100502}
  (\bibinfo{year}{2024}).
\newblock \urlprefix\url{https://doi.org/10.1063/5.0189649}.

\bibitem{Autti2020}
\bibinfo{author}{Autti, S.} \emph{et~al.}
\newblock \bibinfo{title}{{AC} {J}osephson effect between two superfluid time
  crystals}.
\newblock \emph{\bibinfo{journal}{Nat. Mater.}} \textbf{\bibinfo{volume}{20}},
  \bibinfo{pages}{171--174} (\bibinfo{year}{2020}).
\newblock \urlprefix\url{http://dx.doi.org/10.1038/s41563-020-0780-y}.

\bibitem{Kreil2021}
\bibinfo{author}{Kreil, A. J.~E.} \emph{et~al.}
\newblock \bibinfo{title}{Experimental observation of {J}osephson oscillations
  in a room-temperature {B}ose--{E}instein magnon condensate}.
\newblock \emph{\bibinfo{journal}{Phys. Rev. B}}
  \textbf{\bibinfo{volume}{104}}, \bibinfo{pages}{144414}
  (\bibinfo{year}{2021}).
\newblock \urlprefix\url{http://dx.doi.org/10.1103/PhysRevB.104.144414}.

\bibitem{Mohseni2022}
\bibinfo{author}{Mohseni, M.}, \bibinfo{author}{Vasyuchka, V.~I.},
  \bibinfo{author}{L'vov, V.~S.}, \bibinfo{author}{Serga, A.~A.} \&
  \bibinfo{author}{Hillebrands, B.}
\newblock \bibinfo{title}{{Classical analog of qubit logic based on a magnon
  Bose--Einstein condensate}}.
\newblock \emph{\bibinfo{journal}{Commun. Phys.}} \textbf{\bibinfo{volume}{5}},
  \bibinfo{pages}{196} (\bibinfo{year}{2022}).
\newblock \urlprefix\url{https://doi.org/10.1038/s42005-022-00970-8}.

\bibitem{Klaers2010}
\bibinfo{author}{Klaers, J.}, \bibinfo{author}{Schmitt, J.},
  \bibinfo{author}{Vewinger, F.} \& \bibinfo{author}{Weitz, M.}
\newblock \bibinfo{title}{{B}ose--{E}instein condensation of photons in an
  optical microcavity}.
\newblock \emph{\bibinfo{journal}{Nature}} \textbf{\bibinfo{volume}{468}},
  \bibinfo{pages}{545--548} (\bibinfo{year}{2010}).
\newblock \urlprefix\url{http://dx.doi.org/10.1038/nature09567}.

\bibitem{Snoke2002}
\bibinfo{author}{Snoke, D.}
\newblock \bibinfo{title}{Spontaneous {B}ose coherence of excitons and
  polaritons}.
\newblock \emph{\bibinfo{journal}{Science}} \textbf{\bibinfo{volume}{298}},
  \bibinfo{pages}{1368--1372} (\bibinfo{year}{2002}).
\newblock \urlprefix\url{http://dx.doi.org/10.1126/science.1078082}.

\bibitem{Snoke2013}
\bibinfo{author}{Snoke, D.~W.} \& \bibinfo{author}{Girvin, S.~M.}
\newblock \bibinfo{title}{Dynamics of phase coherence onset in {B}ose
  condensates of photons by incoherent phonon emission}.
\newblock \emph{\bibinfo{journal}{J. Low Temp. Phys.}}
  \textbf{\bibinfo{volume}{171}}, \bibinfo{pages}{1--12}
  (\bibinfo{year}{2013}).
\newblock \urlprefix\url{http://dx.doi.org/10.1007/s10909-012-0854-6}.

\bibitem{Frohlich1968}
\bibinfo{author}{Fr{\"o}hlich, H.}
\newblock \bibinfo{title}{Bose condensation of strongly excited longitudinal
  electric modes}.
\newblock \emph{\bibinfo{journal}{Phys. Lett. A}}
  \textbf{\bibinfo{volume}{26}}, \bibinfo{pages}{402--403}
  (\bibinfo{year}{1968}).
\newblock \urlprefix\url{https://dx.doi.org/10/d76sd5}.

\bibitem{Lvov2024}
\bibinfo{author}{L’vov, V.~S.} \emph{et~al.}
\newblock \bibinfo{title}{{B}ose--{E}instein condensation in systems with flux
  equilibrium}.
\newblock \emph{\bibinfo{journal}{Phys. Rev. B}}
  \textbf{\bibinfo{volume}{109}}, \bibinfo{pages}{014301}
  (\bibinfo{year}{2024}).
\newblock \urlprefix\url{http://dx.doi.org/10.1103/PhysRevB.109.014301}.

\bibitem{Schweizer2024}
\bibinfo{author}{Schweizer, M.~R.} \emph{et~al.}
\newblock \bibinfo{title}{Local temperature control of magnon frequency and
  direction of supercurrents in a magnon {B}ose--{E}instein condensate}.
\newblock \emph{\bibinfo{journal}{Appl. Phys. Lett.}}
  \textbf{\bibinfo{volume}{124}}, \bibinfo{pages}{092402}
  (\bibinfo{year}{2024}).
\newblock \urlprefix\url{http://dx.doi.org/10.1063/5.0189154}.

\bibitem{schweizerConfinementBoseEinstein2022}
\bibinfo{author}{Schweizer, M.~R.}, \bibinfo{author}{Kreil, A. J.~E.},
  \bibinfo{author}{{von Freymann}, G.}, \bibinfo{author}{Hillebrands, B.} \&
  \bibinfo{author}{Serga, A.~A.}
\newblock \bibinfo{title}{Confinement of {{Bose}}--{{Einstein}} magnon
  condensates in adjustable complex magnetization landscapes}.
\newblock \emph{\bibinfo{journal}{J. Appl. Phys.}}
  \textbf{\bibinfo{volume}{132}}, \bibinfo{pages}{183908}
  (\bibinfo{year}{2022}).
\newblock \urlprefix\url{https://doi.org/10.1063/5.0123233}.

\bibitem{Bozhko2019}
\bibinfo{author}{Bozhko, D.~A.} \emph{et~al.}
\newblock \bibinfo{title}{Bogoliubov waves and distant transport of magnon
  condensate at room temperature}.
\newblock \emph{\bibinfo{journal}{Nat. Commun.}} \textbf{\bibinfo{volume}{10}},
  \bibinfo{pages}{2460} (\bibinfo{year}{2019}).
\newblock \urlprefix\url{http://dx.doi.org/10.1038/s41467-019-10118-y}.

\bibitem{Kreil2019}
\bibinfo{author}{Kreil, A. J.~E.} \emph{et~al.}
\newblock \bibinfo{title}{Tunable space-time crystal in room-temperature
  magnetodielectrics}.
\newblock \emph{\bibinfo{journal}{Phys. Rev. B}}
  \textbf{\bibinfo{volume}{100}}, \bibinfo{pages}{020406(R)}
  (\bibinfo{year}{2019}).
\newblock \urlprefix\url{http://dx.doi.org/10.1103/PhysRevB.100.020406}.

\bibitem{Demokritov2022}
\bibinfo{author}{Demokritov, S.~O.}
\newblock \bibinfo{title}{Comment on {“Bose--Einstein condensation and spin
  superfluidity of magnons in a perpendicularly magnetized yttrium iron garnet
  film” (JETP Letters 112, 299 (2020))}}.
\newblock \emph{\bibinfo{journal}{JETP Lett.}} \textbf{\bibinfo{volume}{115}},
  \bibinfo{pages}{691--693} (\bibinfo{year}{2022}).
\newblock \urlprefix\url{http://dx.doi.org/10.1134/S0021364022600719}.

\bibitem{Pirro2021}
\bibinfo{author}{Pirro, P.}, \bibinfo{author}{Vasyuchka, V.~I.},
  \bibinfo{author}{Serga, A.~A.} \& \bibinfo{author}{Hillebrands, B.}
\newblock \bibinfo{title}{Advances in coherent magnonics}.
\newblock \emph{\bibinfo{journal}{Nat. Rev. Mater.}}
  \textbf{\bibinfo{volume}{6}}, \bibinfo{pages}{1114--1135}
  (\bibinfo{year}{2021}).
\newblock \urlprefix\url{https://doi.org/10.1038/s41578-021-00332-w}.

\bibitem{Rezende2020}
\bibinfo{author}{Rezende, S.~M.}
\newblock \emph{\bibinfo{title}{Fundamentals of {{Magnonics}}}} Vol.
  \bibinfo{volume}{969} (\bibinfo{publisher}{Springer International
  Publishing}, \bibinfo{address}{Cham}, \bibinfo{year}{2020}).
\newblock \urlprefix\url{https://doi.org/10.1007/978-3-030-41317-0}.

\bibitem{Holstein1940}
\bibinfo{author}{Holstein, T.} \& \bibinfo{author}{Primakoff, H.}
\newblock \bibinfo{title}{Field dependence of the intrinsic domain
  magnetization of a ferromagnet}.
\newblock \emph{\bibinfo{journal}{Phys. Rev.}} \textbf{\bibinfo{volume}{58}},
  \bibinfo{pages}{1098--1113} (\bibinfo{year}{1940}).
\newblock \urlprefix\url{https://link.aps.org/doi/10.1103/PhysRev.58.1098}.

\bibitem{Demidov2008}
\bibinfo{author}{Demidov, V.~E.}, \bibinfo{author}{Dzyapko, O.},
  \bibinfo{author}{Demokritov, S.~O.}, \bibinfo{author}{Melkov, G.~A.} \&
  \bibinfo{author}{Slavin, A.~N.}
\newblock \bibinfo{title}{Observation of spontaneous coherence in
  {B}ose--{E}instein condensate of magnons}.
\newblock \emph{\bibinfo{journal}{Phys. Rev. Lett.}}
  \textbf{\bibinfo{volume}{100}}, \bibinfo{pages}{047205}
  (\bibinfo{year}{2008}).
\newblock
  \urlprefix\url{https://link.aps.org/doi/10.1103/PhysRevLett.100.047205}.

\bibitem{Pethick2008}
\bibinfo{author}{Pethick, C.~J.} \& \bibinfo{author}{Smith, H.}
\newblock \emph{\bibinfo{title}{Bose--{{Einstein Condensation}} in {{Dilute
  Gases}}}}  (\bibinfo{publisher}{Cambridge University Press},
  \bibinfo{year}{2008}).
\newblock \urlprefix\url{https://doi.org/10.1017/CBO9780511802850}.

\bibitem{Serga2014}
\bibinfo{author}{Serga, A.~A.} \emph{et~al.}
\newblock \bibinfo{title}{{B}ose--{E}instein condensation in an ultra-hot gas
  of pumped magnons}.
\newblock \emph{\bibinfo{journal}{Nat. Commun.}} \textbf{\bibinfo{volume}{5}},
  \bibinfo{pages}{3452} (\bibinfo{year}{2014}).
\newblock \urlprefix\url{https://doi.org/10.1038/ncomms4452}.

\bibitem{Schneider2020}
\bibinfo{author}{Schneider, M.} \emph{et~al.}
\newblock \bibinfo{title}{Bose--{{Einstein}} condensation of quasiparticles by
  rapid cooling}.
\newblock \emph{\bibinfo{journal}{Nat. Nanotechnol.}}
  \textbf{\bibinfo{volume}{15}}, \bibinfo{pages}{457--461}
  (\bibinfo{year}{2020}).

\bibitem{Verba2019}
\bibinfo{author}{Verba, R.}, \bibinfo{author}{Tiberkevich, V.} \&
  \bibinfo{author}{Slavin, A.}
\newblock \bibinfo{title}{Hamiltonian formalism for nonlinear spin wave
  dynamics under antisymmetric interactions: Application to
  {D}zyaloshinskii--{M}oriya interaction}.
\newblock \emph{\bibinfo{journal}{Phys. Rev. B}} \textbf{\bibinfo{volume}{99}},
  \bibinfo{pages}{174431} (\bibinfo{year}{2019}).
\newblock \urlprefix\url{https://link.aps.org/doi/10.1103/PhysRevB.99.174431}.

\bibitem{numpy}
\bibinfo{author}{Harris, C.~R.} \emph{et~al.}
\newblock \bibinfo{title}{Array programming with {NumPy}}.
\newblock \emph{\bibinfo{journal}{Nature}} \textbf{\bibinfo{volume}{585}},
  \bibinfo{pages}{357--362} (\bibinfo{year}{2020}).
\newblock \urlprefix\url{https://doi.org/10.1038/s41586-020-2649-2}.

\bibitem{Noack2021}
\bibinfo{author}{Noack, T.~B.} \emph{et~al.}
\newblock \bibinfo{title}{Evolution of room-temperature magnon gas: Toward a
  coherent {B}ose--{E}instein condensate}.
\newblock \emph{\bibinfo{journal}{Phys. Rev. B}}
  \textbf{\bibinfo{volume}{104}}, \bibinfo{pages}{L100410}
  (\bibinfo{year}{2021}).
\newblock
  \urlprefix\url{https://link.aps.org/doi/10.1103/PhysRevB.104.L100410}.

\end{thebibliography}
\end{document}